# Recurrent Neural Networks to automate Quality assessment of Software Requirements

Gramajo María Guadalupe, Ballejos Luciana, Ale Mariel
CIDISI – UTN- FRSF
Lavaisse 610 , Santa Fe, Argentina
{mgramajo, lballejo, male}@frsf.utn.edu.ar

**Abstract**. Many problems related to the quality of requirements arise during elicitation and specification activities, since they are written in natural language. The flexibility and inherent nature of language makes requirements prone to inconsistencies, redundancies, and ambiguities and, consequently, this influences negatively the later phases of the software life cycle. To address this problem, this paper proposes an innovative approach that combines natural language processing techniques and recurrent neural networks to automatically assess the quality of software requirements. Initially, the analysis of singular, complete, correct, and appropriate quality properties defined in the IEEE 29148: 2018 standard is addressed. The proposed neural models are trained with a data set composed of 1000 software requirements. The proposal provides an average accuracy of 75%. These promising results were a motivation to explore its application in the evaluation of other quality properties.

**Keywords.** Software Requirements - Deep Learning - Quality Requirements - Neural Networks

## 1 Introduction

Software requirements are descriptions that reflect the stakeholders' needs regarding what the system should do, the service it offers, and the constraints on its operation (Sommerville, 2011). The field focused on identifying, analyzing, specifying, and validating the requirements of the system to be developed is Requirements Engineering (Abran & Fairley, 2014). Requirements elicitation is a complex and iterative process in which engineers must face the double challenge of discovering and formalizing the stakeholders' needs(Génova et al., 2013)

The success of this process requires intense collaboration and communication with stakeholders throughout the software life cycle (Kummler et al., 2018). Especially, during the early phases of the project, where software requirements are expressed and defined. The need to reduce communication gaps between stakeholders has risen natural language as the most frequently used technique to express software requirements.

Natural language is widely used to acquire and document software requirements, and its use is predominant over other techniques (Dalpiaz et al., 2018; Denger et al., 2003; Femmer, 2013). Although often complemented by models and notations with different degrees of formalization and detail, even nowadays in most software industries requirements are specified using natural language. This trend appears due to the flexibility and comprehensibility provided by natural language, which features include expressivity, intuitivity, and universal means of communication (Sommerville, 2011). These aspects support its use to communicate and document the stakeholders' needs, in a practical and fast way (Kocerka et al., 2018). However, several problems related to the quality of requirements arise, because of the flexibility and inherent nature of language. Unfortunately, its use can lead to inconsistencies, redundancies, and ambiguities around the documentation of software requirements, which negatively influences the subsequent phases of the software life cycle (Denger et al., 2003). Moreover, errors and defects that arise from the low-quality requirements are considered the most expensive and difficult to correct. The main reason is that the fixing work also includes changes in subsequent phases and leads to re-work.

For this reason, it is necessary to provide Requirements Engineering with quality control mechanisms from the initial stages of the project, since, if the requirements are not demanded to meet certain quality criteria in early phases, it will be difficult to find quality in later phases of development. Therefore, it will be even more difficult to obtain a quality software product that provides the features expected by the client. In this context, several authors discuss that achieving *quality software requirements is the first step towards software quality*. In fact, they indicate that the success of a software project is closely related to the quality of its requirements (Curtis et al., 1988; Fabbrini et al., 2001; Kamata & Tetsuo Tamai, 2007, 2007; Knauss & Boustani, 2008).

In order to address this issue, the software community has established guides and standards that define directives to elaborate the Software Requirements Specification (SRS) document and define the requirements included in this artifact, considering quality criteria (IEEE, 1998; INCOSE, 2017; ISO/IEC/IEEE, 2018). These standards and guidelines support the verification and validation activities. In requirement engineer, these include conducting technical reviews to identify and correct inconsistencies, omissions and errors, so that the requirements meet established standards.

However, quality controls are still generally carried out manually, which is not a trivial task and consumes time and effort. This task, besides being considered expensive and error-prone due to its manual condition, generates long feedback loops (Femmer et al., 2016). When a requirements engineer requests a manual review, this implies that multiple stakeholders must be coordinated, including the organization of meetings, consensus and definition of acceptance criteria, discussion of the results and verification of corrected defects. Consequently, this slows down the software development process. Moreover, if the software project expands in size and scope, and with it, the volume of requirements to be reviewed. For these reasons, it is necessary to provide methods and tools that benefit the requirements analysis, in order to produce quality software with controlled time and costs.

In the literature, several proposals can be found that address the analysis of the quality of requirements, (Kummler et al., 2018) point out three main categories. The first category includes those tools developed to assist the engineer in writing the requirements, indicating weaknesses, and giving indications about how the requirements can be improved (Fabbrini et al., 2001; Fantechi et al., 2003; Génova et al., 2013; Wilson et al., 1997). The second category includes studies focused on transforming textual requirements into formal models and logic specifications (Arellano et al., 2015; Holtmann et al., 2011; Ilieva & Ormandjieva, 2005; Verma & Kass, 2008). The third category focuses on the use of machine learning in order to automatically classify requirements according to quality (Huertas & Juárez-Ramírez, 2013; Ormandjieva et al., 2007; Parra et al., 2015; Yang et al., 2012) .

Mainly those proposals that refer to the first two categories mentioned, coincide with the use of natural language processing techniques to first obtain representative sequences of the requirements and then process them through the execution of algorithms that include conditional sentences. These rules are defined by an expert that acts as control mechanisms and allow detecting or restricting the use of regular expressions, grammatical rules, word corpus, and/or previously defined phrases, and thus obtaining quantitative quality indicators and indexes .

However, this type of rule-based strategy leads to evaluation processes that are insufficient and inflexible to evaluate the quality of the requirements, also indicated as arbitrary in the parameterization of the measurements and rigid in the combination of metrics to evaluate the different properties (Moreno et al., 2020). This is because the quality evaluation is limited only to the detection of keywords and/or word corpus, and previously defined grammatical rules, which existence in a requirement or in a requirement document is associated with a penalty value, which is then used to calculate a set of metrics and thus determine the quality. These control mechanisms do not contemplate possible exceptions and semantic and/or syntactic considerations on which a requirements engineer can reflect, when evaluating a requirement in a certain domain, project, or organization. That is the reason why the integration of new technologies to the Requirements Engineering is essential, in order to enrich traditional techniques and procedures that contribute to the quality evaluation.

In relation to this, during the last few years, Artificial Intelligence (AI) has positioned itself as a powerful and accessible tool, capable of being used as a key component in the development of software systems. This is due to the emergence of new learning strategies, availability of large data sets, and increased processing power in computers. The integration of AI technologies in Software Engineering aims at optimizing the development process of software products and automating tasks that demand intensive efforts, in order to obtain intelligent systems with high quality (Gramajo et al., 2020).

However, many proposals that employ machine learning techniques to address the quality of requirements, despite obtaining good results, are characterized by the use of private data sources that prevent repeatable, verifiable, refutable, and/or improvable predictive models. This is essential for the maturity of this research discipline applied to Requirements Engineering.

In this context, the application of AI-derived learning methods and techniques, such as Deep Learning, in tasks that require natural language processing has increased exponentially. This success is mainly due to the fact that deep learning follows the nested hierarchy mechanism for data representation and simulates the human brain. Deep learning can take advantage of processing a large amount of data and achieve results with higher accuracy rates than traditional classification techniques (Alshamrani & Ma, 2019).

In particular, the application of recurrent neural networks (RNNs) has been predominant, due to their capacity to process sequential information. RNNs are considered adequate to model the context dependencies present in language and tasks related to sequence modeling. This feature is very important considering that the words in a language develop their semantics based on the words sequences in the sentences. Another feature that asset its use is the ability to model variable text length (Young et al., 2017). Thus, RNNs are adequate to address the processing and analysis of software requirements expressed in natural language for the study of quality.

This paper presents a set of deep learning models to address the quality of the requirements expressed in natural language, taking as reference four properties defined in the standard IEEE 29148:2018. The quality properties included in this study are *complete*, *appropriate*, *correct*, and *singular*. This proposal aims to provide automatic control mechanisms and quality verification of software requirements. The rest of the paper is organized as follows. Section 2 presents the background. Section 3 presents the methodology and the proposed deep learning models. Section 4 presents materials and performance metrics. Section 5 results and discussions. Finally, the section 6 conclusions and future lines of research.

## 2 Background

This section introduces the concept of the quality properties addressed in this proposal. Then,

the main concepts related to recurrent neural networks are described. LSTM and GRU neural networks are also briefly described.

## Appropriate Quality Property

The specific intent and amount of detail of the requirement is appropriate to the level (level of abstraction) of the entity to which it refers (ISO/IEC/IEEE, 2018). This refers to the fact that a software requirement meets the appropriate quality property if it is expressed in an adequate and consistent level of detail, without specifying how it should be developed. The requirement should not be more detailed than necessary for the level at which it is stated.

## Correct Quality Property

The requirement must be an accurate representation of the entity need from which it was transformed. Therefore, it must be possible to demonstrate that compliance with the requirement, as written, will result in the satisfaction of the need(s) from which it was transformed (ISO/IEC/IEEE, 2018). In other words, a software requirement meets the correct quality property if its specification is an accurate representation of the required need considering the underlying objectives and goals.

## Complete Quality Property

The requirement sufficiently describes the necessary capability, characteristic, constraint, or quality factor to meet the entity need without needing other information to understand the requirement (ISO/IEC/IEEE, 2018).

## Singular Quality Property

The requirement should state a single capability, characteristic, constraint, or quality factor. Although a single requirement consists of a single function, quality or constraint, it can have multiple conditions under which the requirement is to be met (ISO/IEC/IEEE, 2018).

## Recurrent Neural Networks

Since RNNs perform sequential processing by modeling units in sequence, they have the ability to capture the inherent sequential nature of language, where the units are the characters, words, or even phrases (Young et al., 2017). This is very important, because words in a language develop their semantical meaning based on the previous words in the sentence. Moreover, RNNs allow the modeling of such context dependencies in language, which turned out to be a strong motivation for researchers to use RNNs in tasks related to natural language processing. Another feature of this type of neural network is its ability to model variable text length.

An RNN is an extension of a conventional feedforward neural network, which is able to handle a variable-length sequence input. The RNN handles the variable-length sequence by having a recurrent hidden state whose activation at each time is dependent on that of the previous time. Chung et al., 2014 provide a formal definition of RNNs as mentioned below: Given a sequence, $x = (x_1, x_2, ..., x_T)$ the RNN updates its recurrent hidden state $h_t$ by

$$h_t = \begin{cases} 0 & t = 0 \\ \phi(h_{t-1}, x_t), & otherwise \end{cases} \quad (1)$$

where $\Phi$ is a nonlinear function such as composition of a logistic sigmoid with an affine transformation. Optionally, the RNN may have an output $y = (y_1, y_2, ..., y_T)$ which may again be of variable length.

The update of the recurrent hidden state in Eq. (1) is implemented as

$$h_t = g(Wx_t + Uh_{t-1}) \quad (2)$$

where $g$ is a smooth, bounded function such as a logistic sigmoid function or a hyperbolic tangent function. A generative RNN outputs a probability distribution over the next element of the sequence, given its current state $h_t$ and this generative model can capture a distribution over sequences of variable length by using a special output symbol to represent the end of the sequence. The sequence probability can be decomposed into

$$p(x_1, ..., x_T) = p(x_1)\, p(x_2|x_1)\, p(x_3|x_1, x_2) ... p(x_T|x_1, ..., x_{T-1}) \quad (3)$$

where the last element is a special end-of-sequence value. We model each conditional probability distribution with

$$p(x_t x_1, ..., x_{t-1}) = g(h_t) \quad (4)$$

where $h_t$ is from Eq. (1).

Unfortunately, it has been observed that it is difficult to train RNNs to capture long-term dependencies because the gradients tend to disappear or explode. This limitation was overcome by several networks, such as long term memory (LSTM), gated recurrent units (GRU), and residual networks (ResNets). In this paper, we evaluate the performance of LSTM and GRU since, as the literature indicates, they are the most used in natural language processing applications (Young et al., 2017).

## Long Short Term Memory

LSTM has an additional "forget" gate over the simple RNN. Its unique mechanism enables it to overcome both the vanishing and exploding gradient problem (Gers et al., 2000; Hochreiter & Schmidhuber, 1997).

Unlike the vanilla RNN, LSTM allows the error to back-propagate through an unlimited number of time steps. Consisting of three gates: input, forget and output gates, it calculates the hidden state by taking a combination of these three gates as per the equations below:

$$x = \begin{bmatrix} h_{t-1} \\ x_t \end{bmatrix} \quad (5)$$

$$f_t = \sigma(W_f x + b_f) \quad (6)$$

$$i_t = \sigma(W_i x + b_i) \quad (7)$$

$$o_t = \sigma(W_o x + b_o) \quad (8)$$

$$c_t = f_t \odot c_{t-1} + i_t \odot tanh(W_c X + b_c) \quad (9)$$

$$h_t = o_t \odot tanh(c_t) \quad (10)$$

### Gated Recurrent Unit

The GRU consists of two gates, the reset gate, and the update gate, and handles the flow of information as an LSTM without a memory unit. Therefore, it exposes all the hidden content without any control. The formulas of GRU are as follows:

$$z = \sigma(U_z x_t + W_z h_{t-1}) \quad (11)$$

$$r = \sigma(U_r x_t + W_r h_{t-1}) \quad (12)$$

$$s_t = tanh(U_z x_t + W_s(h_{t-1} \odot r)) \quad (13)$$

$$h_t = (1-z) \odot s_t + z \odot h_{t-1} \quad (14)$$

## 3 Methodology

The workflow used to develop the proposed deep learning models is based on the guidelines of Amershi et al., 2019. It consists of different phases, some of which are oriented to data management (collection, cleaning, and labeling), and others are oriented to the definition and configuration of the model (requirements, features engineering, training, evaluation, implementation, and monitoring). The flow also includes a series of feedback loops that allow the model to be adjusted. Fig. 1 shows the machine learning workflow. Below are the phases that describe the workflow used.

### Model Requirements

In this stage, the appropriate model is selected for the problem to be solved (classification, regression, clustering, among others) and it is discussed which characteristics are feasible to be analyzed by machine learning techniques.

### Data Collection

In the data collection phase, data sets are collected and merged. This phase is critical and has an absolute influence on the performance of the machine learning model. In this phase, raw data sets are formed. Raw data refers to the fact that they are in their original format, without any previous preparation.

### Data Cleaning

In the data cleaning phase, the noises and anomalies present in the data are removed. From the raw data, corrupted records or invalid values are removed or corrected.

### Data Labeling

Data labeling is essential in the development of machine learning models since for the training of such models, it is necessary to provide data samples representative of what you want to predict. In this phase, each of the records contained in the data collection is tagged, that is, each record is assigned a category according to what you want the model to learn.

### Feature Engineering

The feature engineering phase involves the activities related in which variables are extracted and selected or augmented as needed for the machine learning models. The feature engineering is fundamental since it allows to define the inputs for model training with the appropriate characteristics and format. In this phase, adjustments are also made to the data to create the attributes required by the machine learning model. Examples of such adjustments are the numerical scaling of the columns of the data set to a value between 0 and 1, the clipping of values, and the categorical one-hot coding characteristics, which constitute the set of pre-processing and transformation operations.

### Model Training

In the model training phase, as its name suggests, the model is trained with the pre-processed data. Such data is already in the appropriate input format for the model. In this phase the model is fed with data which value for the target data attribute (characteristic to be predicted) is known but not included. The model is then run to predict the target values of the training data. A common strategy is to take all the labeled data and divide it into training and evaluation subsets, usually with a ratio of 70 to 80% for training and 20 to 30% for evaluation.

### Model Evaluation

In this phase, the model is fed with data including the target values. The results of the model predictions are then compared with the actual values of the evaluation data, i.e. the model output is evaluated. It is possible to determine the predictive quality of the model using performance metrics. It is important to note that model evaluation and monitoring may loop back to any of the previous stages, represented in the Fig. 1 with the adjusting nuts, and that model training may loop back to feature engineering.

In this proposal, the phases of model implementation and monitoring in real environments are part of future work.

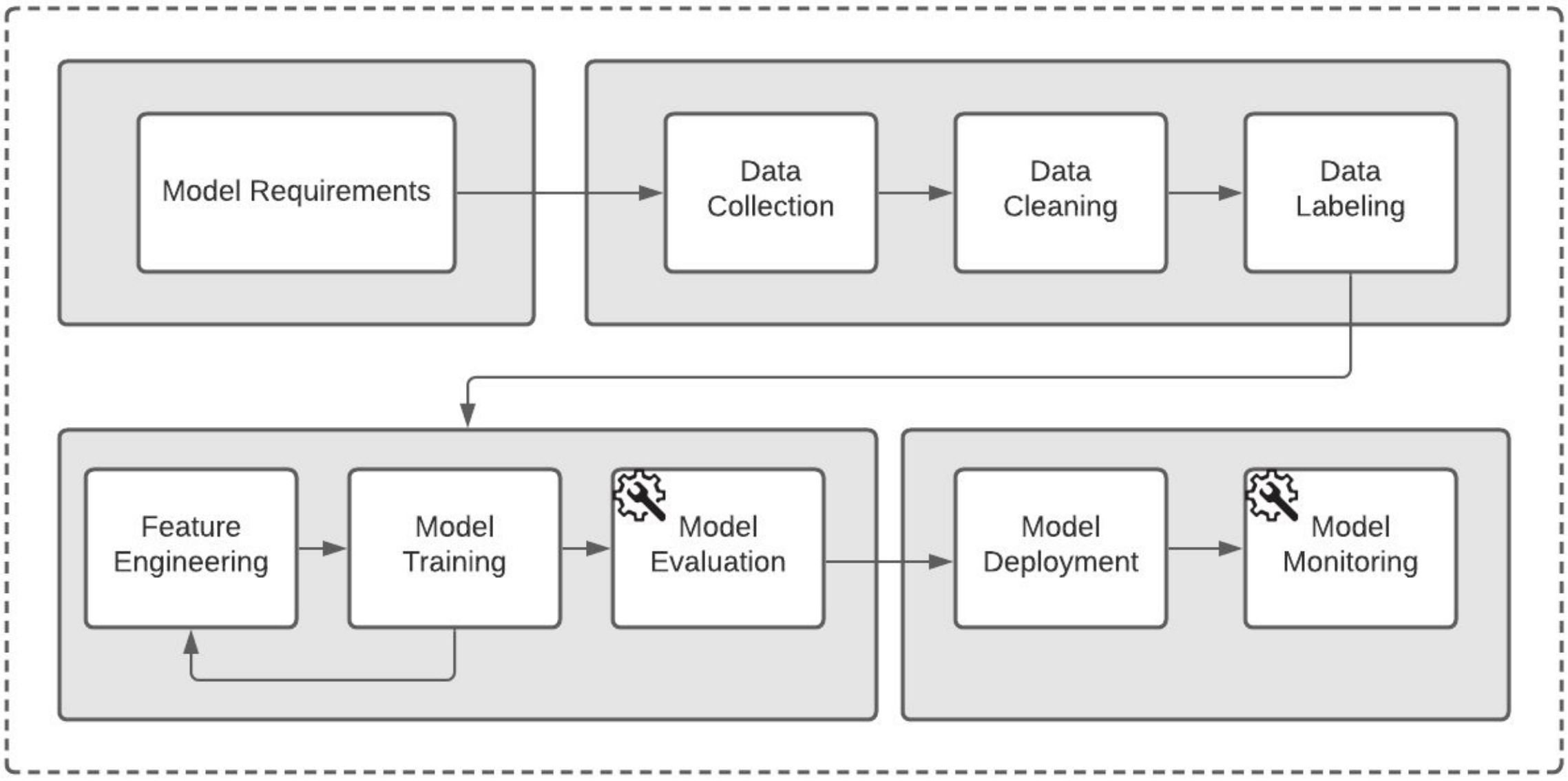

**Fig. 1** Machine Learning Workflow

## Proposal of Deep Learning Models

This paper presents a binary classification model for each quality property under study. The models have a supervised learning approach. For the training of each classification model, the pre-processed data set is vectorized using word embedding. It takes the data set and produces a vector of M dimensions by each requirement. Then, the embedding layer transforms each element (i) of the vectors obtained in the previous step into a vector space of M X N dimensions. Each neural model has an LSTM layer that allows the processing of sequential events. The latter is connected to a fully connected layer with a softmax activation function. The softmax function calculates the probability distribution of each requirement over different classes defined.

The architecture of the binary classification models used by each quality property is shown in Fig. 2.

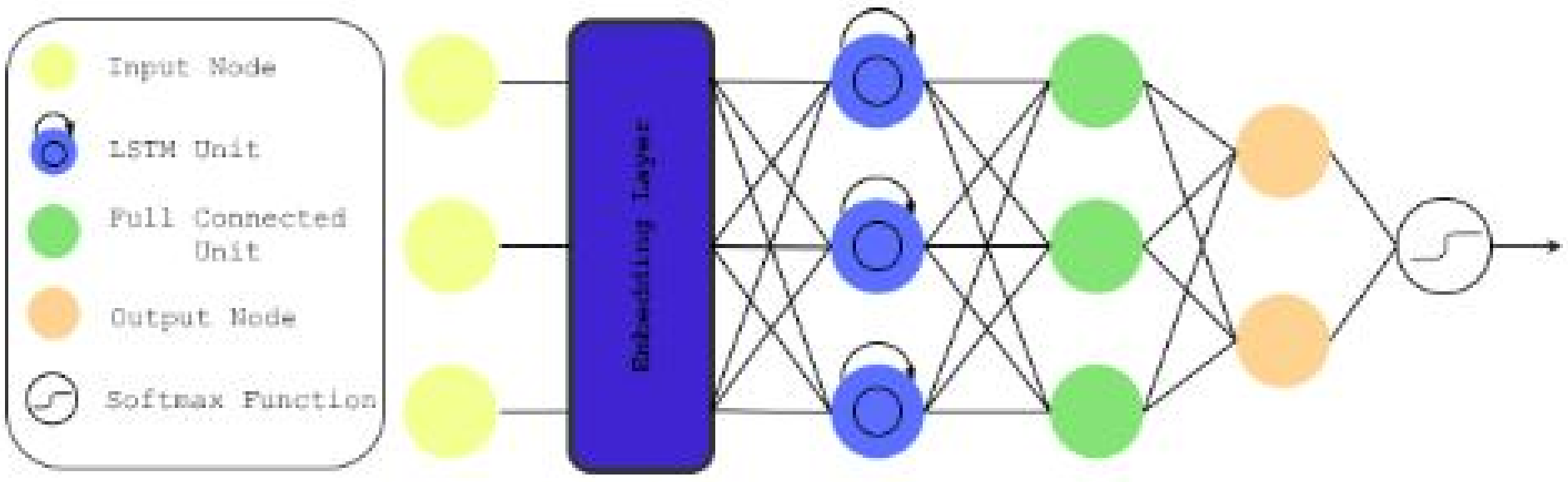

**Fig. 2** Deep Learning Architecture

Each defined neural model includes 40 input units, which correspond to the vocabulary of grammatical tags used at least once in the pre-processing to represent the requirements. These units allow the entry of the requirements contained in a suitable input format to the embedding layer. The output dimension of the embedding layer is 128, which corresponds to the size of the vector space in which the requirements will be embedded.

As mentioned above the proposed architecture for each model also includes an LSTM layer that connects to a full connected layer. Both mentioned layers are composed of three elements: number of input units (num-input), number of output units (num-output), and number of layers (num-layers).

The LSTM layer has 128 input units since it considers the output dimension of the embedded layer. The number of output units of the LSTM layer has a value equal to 128. Meanwhile, the number of layers has a value of 1.

The full connected layer has 128 input units corresponding to the output units of the LSTM layer. Meanwhile, the number of output units has a value of 2, which allows representing the classes defined by each model in this proposal. It should be noted that four binary classification models were defined whose classes correspond to (singular - not singular), (complete - not complete), (appropriate - not appropriate), (correct - not correct), which represent the quality properties of the requirements to be predicted.

The number of layers has a value of 1. In addition, the units in this layer are subject to a dropout probability of 0.3. Finally, as mentioned above, the softmax function determines the probability of success of each defined class. Table 1 summarizes the hyperparameters of the defined models.

Table 1 Hyperparameters

| Hyperparameter | Value |
|---|---|
| Input Dimension Embedding Layer | 40 |
| Output Dimension Embedding Layer | 128 |
| Input LSTM Layer | 128 |
| Output LSTM Layer | 128 |
| Input Full connected Layer | 128 |
| Output Full connected Layer | 2 |
| Dropout | 0.3 |

## Experimental setup

To avoid overfitting, 10-fold cross-validation has been made for each classification model. In each fold, we have used the training and test partitions, 80% for training, and 20% for testing. The model parameters (weights) have been tuned with the Adam optimizer.

The parameters of the models were determined with an inner grid search of a range of possible values, to be executed with the data set. The Table 2 summarizes the parameters of grid search defined. This technique allows executing each neuronal model with a set of random configurations, which facilitates later, to identify the configurations that return the best performance. It should be noted that this proposal also considers the second variant of RNNs, Gated Recurrent Unit (GRU), as a parameter for each defined classification model.

Table 2 Parameters

| Parameter | Value |
|---|---|
| Epoch | [3, 4, 5, 10, 30, 40, 100] |
| Learning Rate | [0.1, 0.01, 0.001] |
| Embedding Dimension | [64, 128, 256, 2048] |
| Number Layers | [1, 2] |
| Number Units | [64, 128, 256, 1024] |
| Dropout | [0, 0.1, 0.3] |
| Layer Type | LSTM, GRU |
| Optimizer | Adam |
| Error Criteria | Binary Cross Entropy |

# 4 Materials and Performance Metrics

## Requirements Data set

In order to provide the data to train the defined neural models, a data set composed of 1000 software requirements expressed in natural language extracted from Public Requirements Data set (PURE) was created. The PURE data set provides a collection of 79 documents of requirements written in English extracted from the web. They cover multiple domains from various companies and university projects which contributes to the heterogeneity of the data included in our data set (Ferrari et al., 2017).
The requirements that make up the data set were randomly selected and manually extracted from 26 documents. Noises and anomalies attributable to writing styles, syntax, and spelling were not eliminated since they are necessary for the analysis of the quality perceived by the neural models. Then, the extracted requirements were manually classified by expert requirements engineers. The labels used for the classification were *singular*, *complete*, *appropriate*, and *correct* representing the quality properties under analysis in this paper, defined by the ISO/IEC/IEEE, 2018 standard.

## Data Preprocessing

The requirements contained in the data set were pre-processed in order to structure them into an input format suitable for each neural model. First, the NLTK library tokenizer was used to divide each input requirement into tokens (Stanford NLP Group, 2020b). Tokens can be words, numbers, punctuation marks, or symbols. Then, the Part Of Speech tagging was performed, which consists of associating the tokens that make up each requirement with its corresponding grammatical tag. The label set used is the one provided by Stanford NLP

Group, 2020a. Then, each tagged requirement is transformed into a numerical representation using the word embedding technique.
The variable length of the requirements that make up the data set was established as maximum length ($L$) 100 words per requirement. Taking as a reference the longest requirement included in our data set. For those requirements of less than L length, the pre sequence padding strategy was adopted. This technique consists of filling in the numerical sequences with zeros (from left to right) until the established length is reached. Pre sequence padding is a technique widely used in prediction problems with variable-length input data. The numerical representations obtained are used to feed the deep learning networks, which learn to automatically classify the quality of the requirements. Fig. 3 illustrates the data preprocessing.

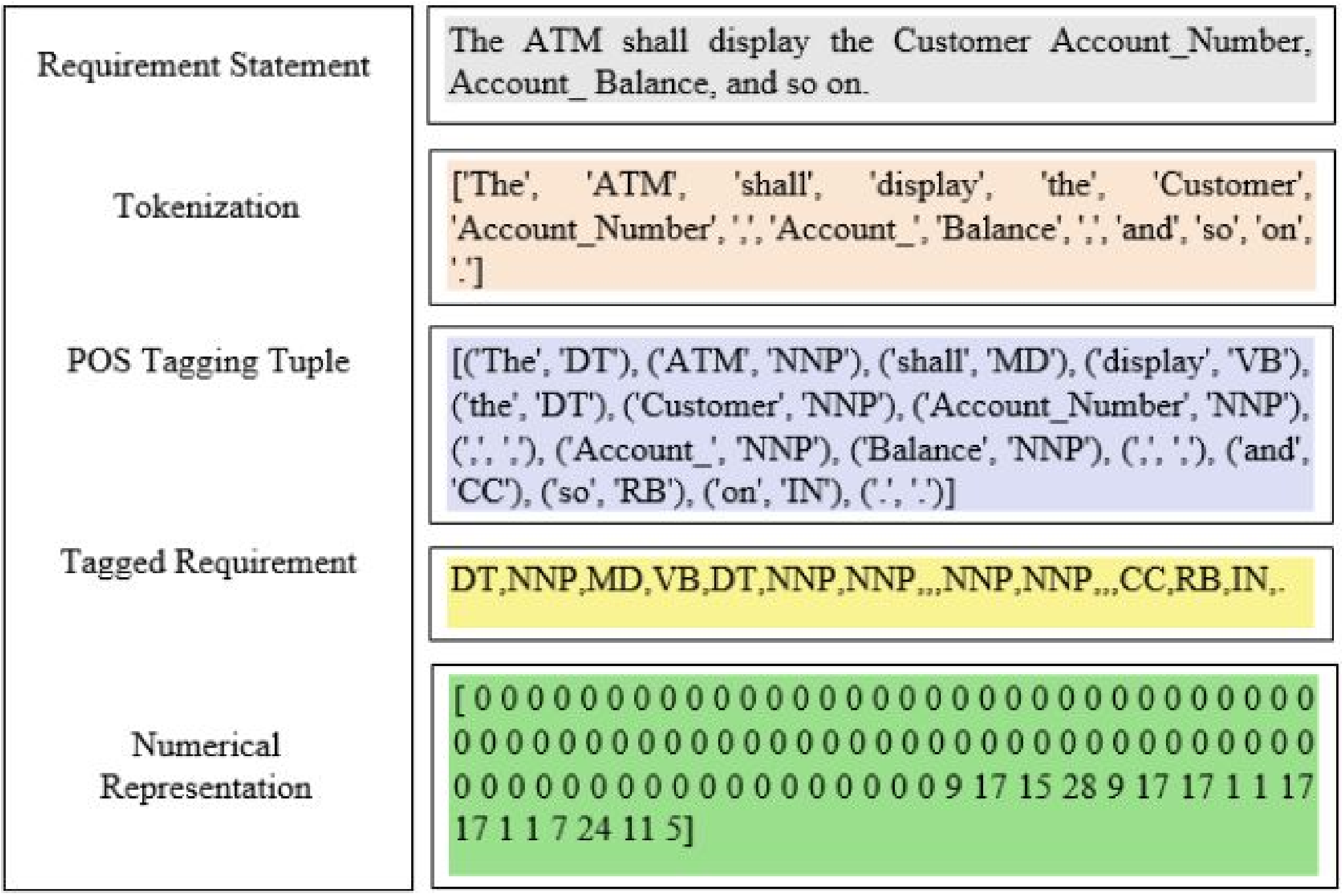

**Fig. 3** Data Preprocessing

## Performance Evaluation

The performance of the proposed classification models was evaluated using the following standard performance metrics: precision ($P$), recall ($R$), accuracy ($A$), and harmonic mean between the sensitivity (+) and precision ($F_1$).

$$P = \frac{TP}{TP+FP} \qquad A = \frac{TP+TN}{P+N}$$

$$R = \frac{TP}{TP+FN} \qquad F_1 = \frac{P \cdot R}{P+R}$$

where *TP*, *TN*, *FP* and *FN* are the number of true positive, true negative, false positive and false negative classifications, respectively.
Recall or sensitivity (+) is the proportion of true positive cases that are correctly predicted positive. In other words, the *R* measures how good a rating method is for recognizing and not losing the true positives. Precision or Confidence denotes the proportion of predicted positive cases that are correctly true positives. Therefore *P* measures the relation between true positives and false positives. Regarding accuracy *A* measures the ratio of correct predictions. Finally, $F_1$, being the harmonic score between *P* and sensitivity (+).
Another aspect to be mentioned is the hardware and software characteristics used in the execution of the neural models. These are, Linux 4.14.137 SMP x86 64; RAM 12GB; Google Compute Engine GPU and Python 3.6.8.

# 5 Results and Discussions

This section presents the results obtained for each defined binary classification model.
The execution of the best configuration obtained for the quality property "*complete*" classification model is shown in Table 3. This table summarizes the values of the parameters found and the values that the model performed with such configuration.

Table 3 Quality Property Complete

| Parameter | Value |
|---|---|
| Learning Rate | 0.01 |
| Epochs | 5 |
| Dropout | 0 |
| Embedding Dimension | 64 |
| Num Layers | 1 |
| Num Units | 256 |
| Layer Type | GRU |
| Precision | 0.75 |
| Accuracy | 0.75 |
| Recall | 1.0 |
| $F_1$ | 0.85 |

The Fig. 4 below shows that the model achieves the highest precision for the quality property "*complete*" in epoch 5. In the latter, the accuracy achieved is 75%. As far as the model is concerned, it has a mean square error (MSE) of 0.49.

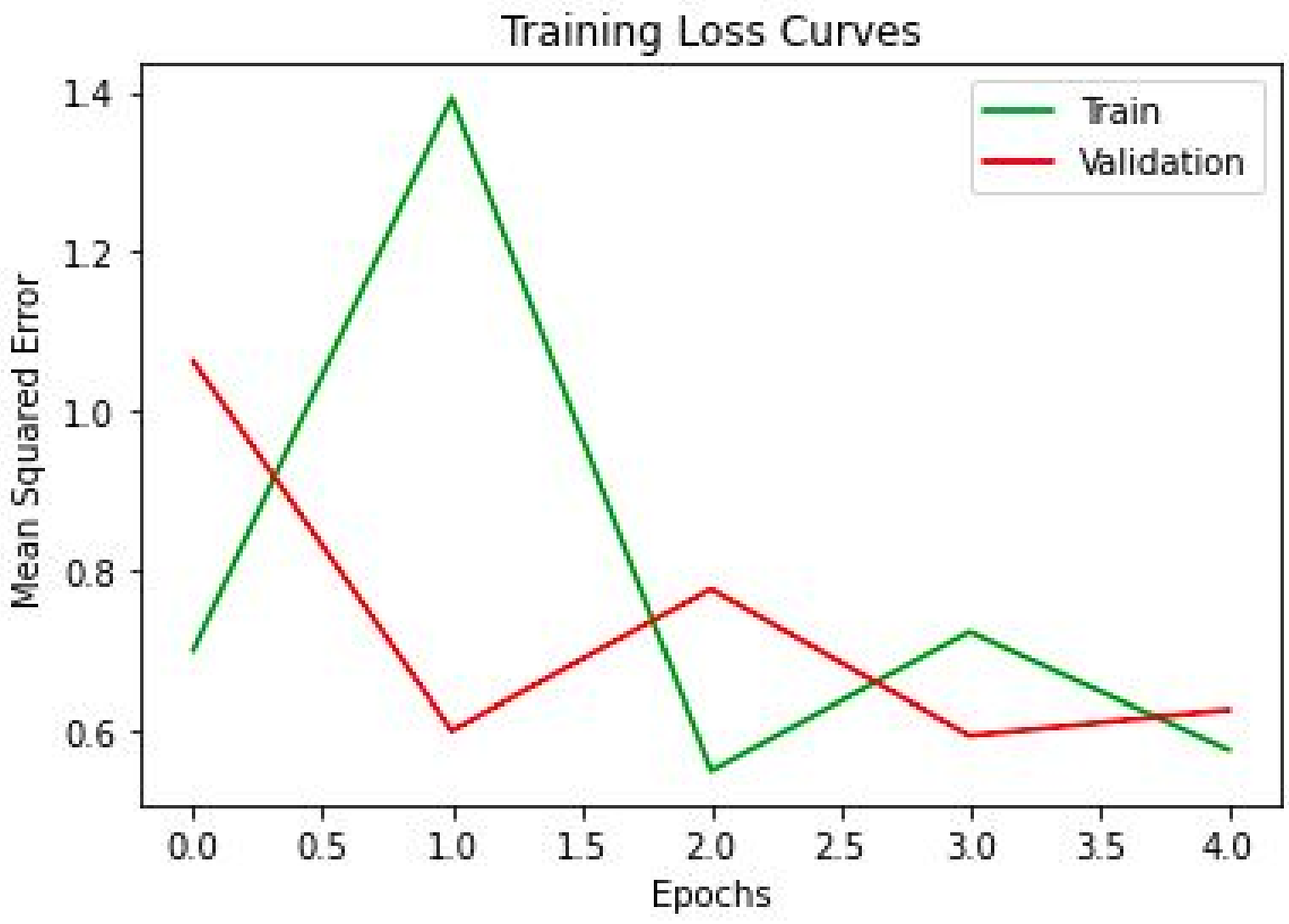

**Fig. 4** Training Loss Curves of Quality Property Complete

Regarding the quality property "singular", the best configuration and the values of the performance measures obtained are shown in Table 4.

Table 4 Quality property Singular

| Parameter | Value |
|---|---|
| Learning Rate | 0.01 |
| Epochs | 40 |
| Dropout | 0.3 |
| Embedding Dimension | 128 |
| Num Layers | 1 |
| Num Units | 64 |
| Layer Type | GRU |
| Precision | 0.78 |
| Accuracy | 0.77 |
| Recall | 0.86 |
| $F_1$ | 0.82 |

The Fig. 5 shows that the model achieves the highest precision for the quality property "*singular*" in epoch 40. In the latter, the accuracy achieved is 78%. As far as the model is concerned, it has a mean square error (MSE) of 0.32.

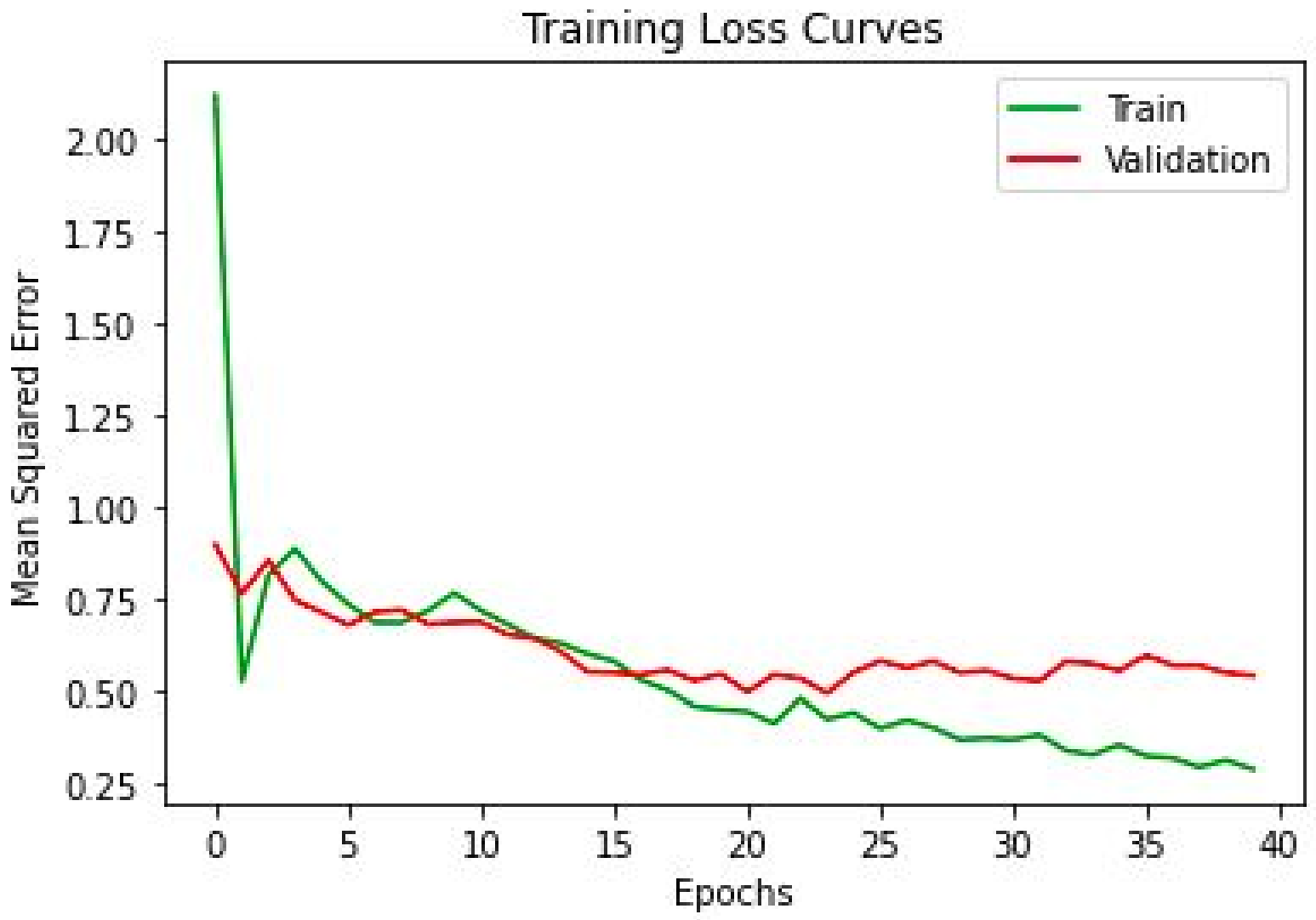

Fig 5. Training Loss Curves of Quality Property Singular

The quality property "*appropriate*" finds its optimal with the following configuration (Table 5).

Table 5 Quality Property Appropriate

| Parameter | Value |
|---|---|
| Learning Rate | 0.001 |
| Epochs | 100 |
| Dropout | 0.3 |
| Embedding Dimension | 2048 |
| Num Layers | 1 |
| Num Units | 1024 |
| Layer Type | GRU |
| Precision | 0.72 |
| Accuracy | 0.70 |
| Recall | 0.82 |
| $F_1$ | 0.76 |

The Fig. 6 shows that the model achieves the highest precision for the quality property "*appropriate*" in epoch 100. In the latter, the accuracy achieved is 72%. As far as the model is concerned, it has a mean square error (MSE) of 0.23.

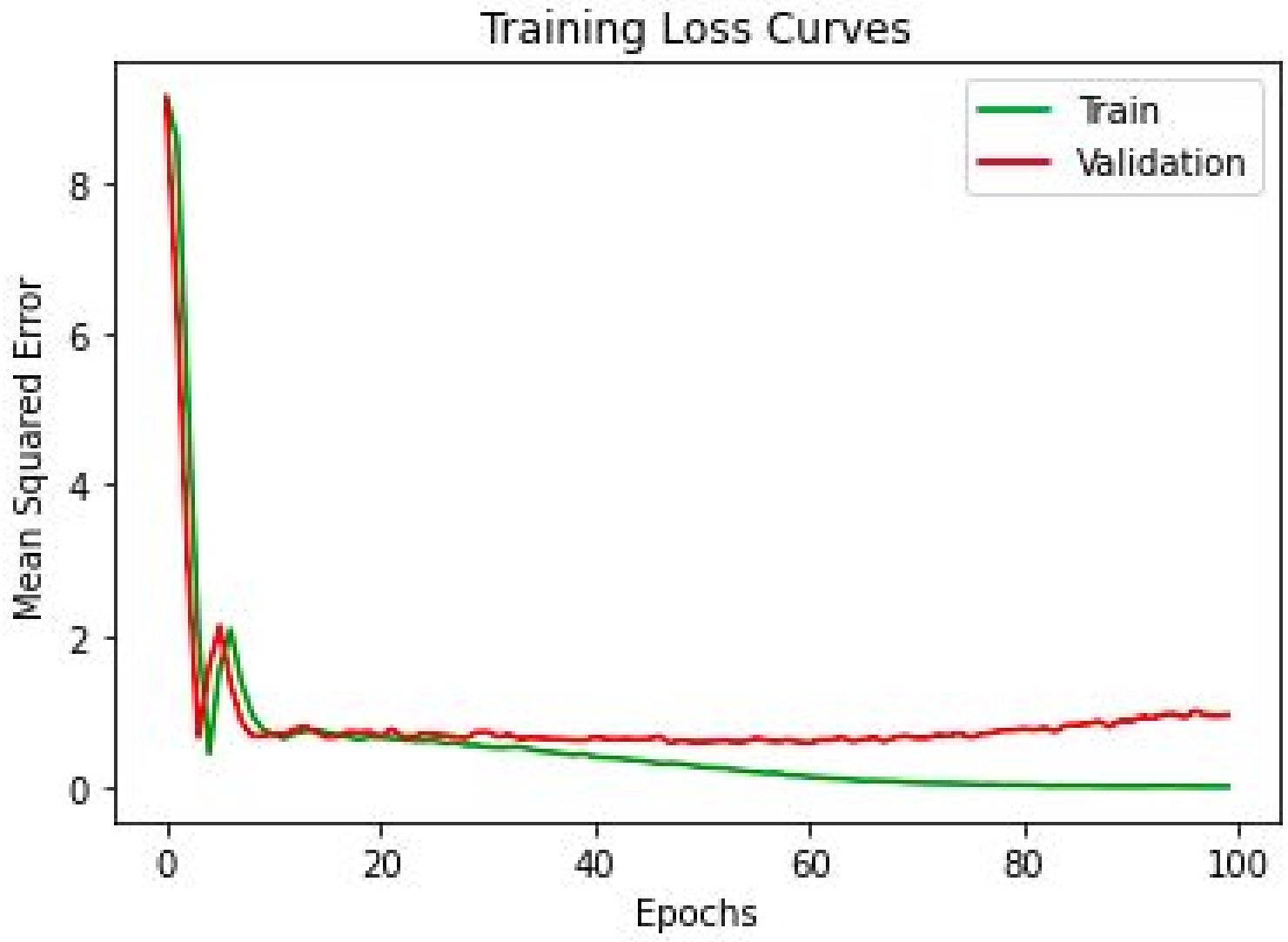

**Fig. 6** Training Loss Curves of Quality Property Appropriate

The best configuration found and their values for the quality property "*correct*" are presented in Table 6.

Table 6 Quality Property Correct

| Parameter | Value |
|---|---|
| Learning Rate | 0.01 |
| Epochs | 4 |
| Dropout | 0 |
| Embedding Dimension | 128 |
| Num Layers | 1 |
| Num Units | 64 |
| Layer Type | GRU |
| Precision | 0.75 |
| Accuracy | 0.75 |
| Recall | 1.0 |
| $F_1$ | 0.85 |

Fig. 7 shows that the model achieves the highest precision for the quality property "*correct*" in epoch 4. In the latter, the accuracy achieved is 75%. As far as the model is concerned, it has a mean square error (MSE) of 0.49.

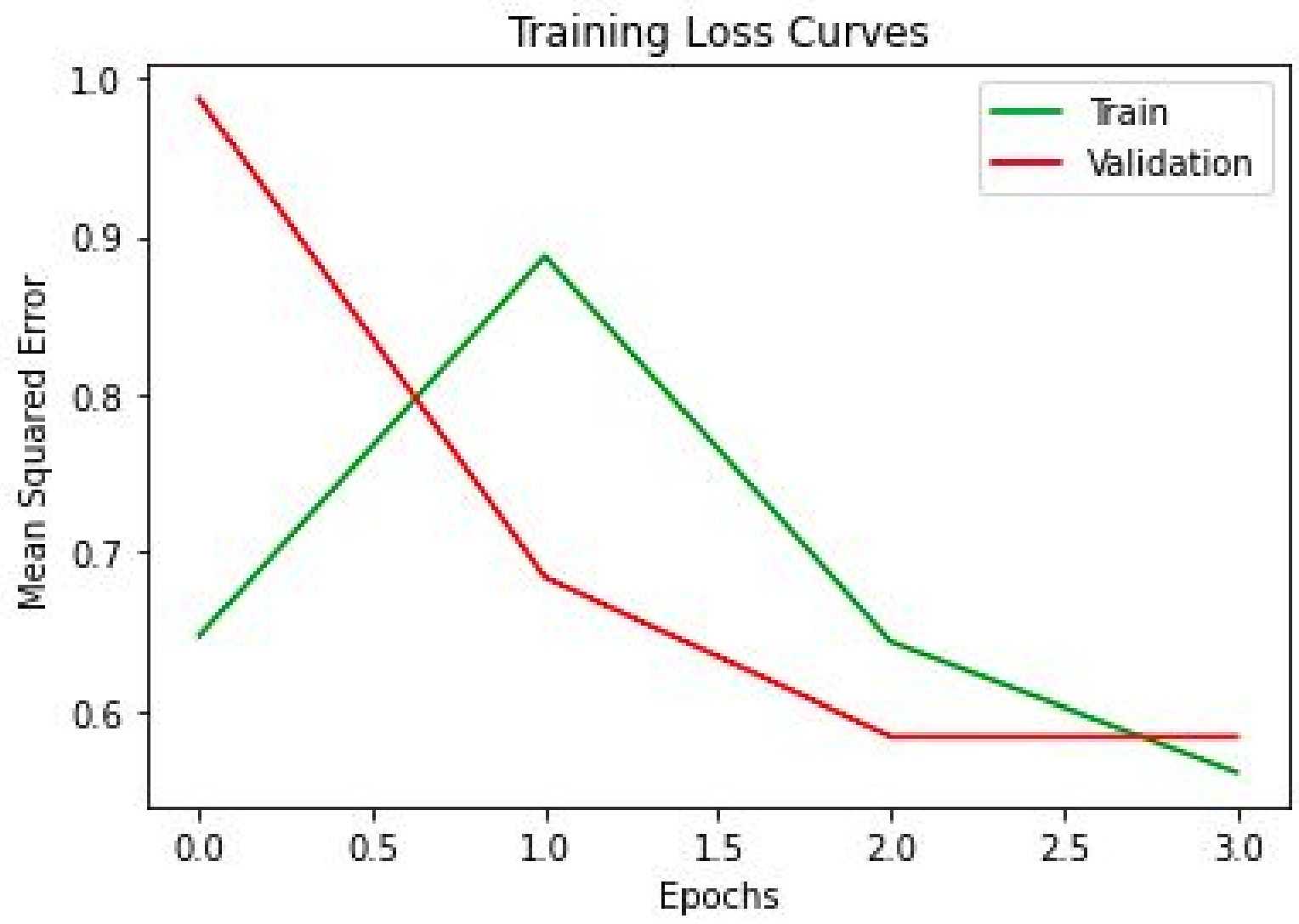

**Fig. 7** Training Loss Curves of Quality Property Correct

From the results obtained, it is possible to determine that this configuration is promising. The average accuracy achieved is 75%. This motivates its exploration in the evaluation of other quality properties, in order to demonstrate that the proposed approach can support the overall quality evaluation of the requirements documented in a Requirements Specification.

Another aspect that is observed from the results is that the GRU's neural networks obtain results of higher precision regarding the LSTM. This can be considered as a starting point for future researches that wish to address the study of the quality of software requirements through the use of recurrent neural networks. The reason is that the authors have not found other research using this type of network in the study of this topic.

## Threats to Validity

This section discusses possible threats to validity with respect to the proposed quality assessment approach.

Since neural models have a supervised learning approach, there are potential threats that could affect their validity. Mainly, this is because the performance is highly dependent on their inputs, i.e. on the labeled requirements by the experts. Despite the authors' efforts to decrease biases in this regard, the labeled requirements used in model training could represent a potential threat to validity. However, to mitigate this, the experts considered the guidelines to write requirements proposed by INCOSE, 2017 during the requirements classification process. These guidelines help engineers to determine whether a requirement meets a particular quality property.

Another limitation that can be observed as a result of the execution of the classification models is the presence of false positives and false negatives. Which existence is associated with the need to enrich the set of requirements used during training.

## 6 Conclusions

This paper introduced a new approach that uses recurrent neural networks to address the assessment of quality properties of software requirements written in natural language. These quality properties are: *singular*; *correct*; *complete;* and *appropriate*. To achieve this goal four binary classification models were defined. The requirements were first pre-processed and structured in an input format suitable for each neuronal model. The proposed binary classification models use a data set composed of 1000 training requirements. The generated data set has been partitioned, considering 80% of the sample to perform the neural network training. Meanwhile, 10% of the sample has been used for testing and the remaining 10% for validation. The cross validation technique is applied in this proposal to avoid model overfitting and the random grid search technique to optimize the parameters.

The results obtained are promising and encourage us to explore its application on other quality properties. Also, it should be noted that the defined neuronal models reach their best performance with GRU over LSTM. On the other hand, although there are other proposals in the literature that use machine learning techniques to address the quality of requirements, the approach presented in this study is novel, because it uses recurrent neural networks. These neural networks allow us to capture the inherent nature of language given its ability to process sequential information. As mentioned in previous sections, this feature is very important, because words in a language develop their semantical meaning based on the previous words in the sentence. Consequently, the training of the classification models is done considering the semantics and syntax of the labeled requirements that compose the data set. It should be noted that we have not identified any other study that used this type of neural networks to address the quality of requirements.

As future work it is intended to apply the proposed approach to the evaluation and analysis of the remaining individual quality properties of the requirements defined in the standard ISO/IEC/IEEE, 2018. And then, to integrate the solution to a framework that allows the evaluation of the quality of the requirements in an automatic way.

## Acknowledgments

The authors would like to thank the following institutions for their support: CONICET y Universidad Tecnológica Nacional (SIUTIFE0004923TC).